\documentclass[twocolumn,preprintnumbers,amsmath,amssymb,aps,physrev,superscriptaddress]{revtex4-2} 

\usepackage{amsmath} 
\usepackage[english]{babel} 
\usepackage{graphicx} 
\usepackage{listings} 
\usepackage{soul}
\usepackage{ulem}
\usepackage{color}
\usepackage[flushleft]{threeparttable}

\begin{document}

\title{Protein-like dynamical transition of hydrated polymer chains}

\author{L.~Tavagnacco}
\affiliation{CNR-ISC and Department of Physics, Sapienza University of Rome, I-00185 Roma, Italy}

\author{M.~Zanatta}
\affiliation{Department of Physics, University of Trento, I-38123 Trento, Italy}

\author{E.~Buratti}
\affiliation{CNR-ISC and Department of Physics, Sapienza University of Rome, I-00185 Roma, Italy}

\author{B.~Rosi}
\affiliation{Department of Physics and Geology, University of Perugia, I-06123, Perugia, Italy}

\author{B.~Frick}
\affiliation{Institut Laue Langevin, F-38042 Grenoble, France}

\author{F.~Natali}
\affiliation{CNR-IOM, Operative Group Grenoble (OGG), Institut Laue Langevin, F-38042 Grenoble, France}

\author{J.~Ollivier}
\affiliation{Institut Laue Langevin, F-38042 Grenoble, France}

\author{E.~Chiessi}
\affiliation{Department of Chemical Science and Technologies, University of Rome Tor Vergata, I-00133 Roma, Italy}

\author{M.~Bertoldo}
\affiliation{Department of Chemical and Pharmaceutical Sciences, University of Ferrara, I-44121 Ferrara, Italy}
\affiliation{CNR-ISOF, I-40129 Bologna, Italy}

\author{E.~Zaccarelli}
\email[Corresponding author: ]{emanuela.zaccarelli@cnr.it}
\affiliation{CNR-ISC and Department of Physics, Sapienza University of Rome, I-00185 Roma, Italy}

\author{A.~Orecchini}
\email[Corresponding author: ]{andrea.orecchini@unipg.it}
\affiliation{Department of Physics and Geology, University of Perugia, I-06123, Perugia, Italy}
\affiliation{CNR-IOM c/o Department of Physics and Geology, University of Perugia, I-06123, Perugia, Italy}
\date{\today}

\begin{abstract}
Combining elastic incoherent neutron scattering experiments at different resolutions with molecular dynamics simulations, we report the observation of a protein-like dynamical transition in linear chains of Poly(N-isopropylacrylamide). We identify the onset of the transition at a temperature $T_d$ of about 225~K. Thanks to a novel global fit procedure, we find quantitative agreement between measured and calculated polymer mean-squared displacements at all temperatures and time resolutions. Our results confirm the generality of the dynamical transition in macromolecular systems in aqueous environments, independently of the internal polymer topology.
\end{abstract}

\maketitle

\section{Introduction}
\label{sec:intro}
A long-debated phenomenon in the (bio)physical community is the occurrence of a dynamical transition in proteins, either globular or intrinsically disordered, that has been widely investigated by means of neutron scattering experiments~\cite{doster1989dynamical}. As its name suggests, such a transition appears as a change of the protein dynamical properties, namely a steep increase of their atomic mobility, due to the onset of anharmonic motions, at a temperature $T_d$ whose value depends on the specific system. A dynamical transition was observed at $T_d$ between 180 K and 230 K for proteins such as myoglobin and lysozyme~\cite{doster1989dynamical,capaccioli2013dependence,Schiro2019rev}, while values up to $\simeq$240 K and $\simeq$260 K were detected for intrinsically disordered proteins~\cite{schiro2015translational} and in purple membranes~\cite{wood2007coupling}, respectively. At the same $T_d$, the atomic motions that are responsible for protein biological function at physiological temperatures start to become active~\cite{Ferrand1993,Zaccai2000}. Much of the debate has been focused on the role played by water in the transition, with several works supporting a water-induced scenario~\cite{Tarek2002,Wood2008,nakagawa2010percolation,schiro2015translational,Sterpone2017,Schiro2019rev}, but also with recent findings of a dynamical transition in dry protein powders~\cite{liu2017dynamical}. Another direction of investigation concerns the generality of the phenomenon, that was found to occur in different bio-macromolecules, including DNA~\cite{Cornicchi2007}, RNA~\cite{Caliskan2006} and lipid bilayers~\cite{Peters2017}. Along this line, recent works~\cite{zanatta2018evidence,tavagnacco2019water} reported evidence of a dynamical transition in concentrated microgel suspensions of Poly(N-isopropylacrylamide) (PNIPAM), extending the realm of the dynamical transition to non-biological systems. Microgels are colloidal-scale particles made by covalently cross-linked polymer networks~\cite{fernandez2011microgel}, that were shown to avoid water crystallization at low temperatures even in highly hydrated samples, i.e. down to a polymer weight fraction of $\sim43$~wt\%~\cite{zanatta2018evidence}. For these systems, a clear increase of the experimental mean-squared displacement (MSD) of the polymer atoms was observed, whereas atomistic molecular dynamics simulations highlighted the pivotal role of water in the transition~\cite{tavagnacco2019water}.

In the biophysical context, a dynamical transition was observed for structured proteins and also for their building blocks, namely polypeptides~\cite{he2008protein} and even amino acids~\cite{schiro2011protein}. Following a similar direction, it is now legitimate to ask whether the polymeric architecture has any influence on the occurrence of such a transition by examining the case of PNIPAM linear (non cross-linked) polymer chains. This polymer is mostly exploited for its thermoresponsive properties~\cite{rubio2008,karg2019nanogels,rovigatti2019numerical}: above room temperature, PNIPAM chains undergo a reversible coil-to-globule transition with increasing temperature $T$, which makes them suitable to mimic protein folding and to investigate protein cold denaturation~\cite{Ando1989}. The similarity of PNIPAM with proteins can be traced back to its amphiphilic character and the associated complex energy landscape, endowed with multiple conformational sub-states of close energy~\cite{tiktopulo1995domain}. Notwithstanding the wide literature on the solution behavior of PNIPAM chains, very little is known on their properties at high concentrations and in the low-temperature regime~\cite{Afroze2000}.

In this paper, we provide a comprehensive investigation of the atomic dynamics of PNIPAM linear chains at low $T$, combining elastic incoherent neutron scattering (EINS) experiments at various energy resolutions and atomistic molecular dynamics (MD) simulations. Our results show the occurrence of a dynamical transition at $T_d \sim 225$~K, a value very similar to that observed in proteins such as myoglobin and lysozyme~\cite{doster1989dynamical,capaccioli2013dependence,Schiro2019rev}. This result implies a wide generality of the phenomenon, independently of the structural details of the investigated complex macromolecular system.

\section{Experimental and numerical methods}
\label{sec:materials}

\subsection{Sample preparation}
\begin{figure}[htbp]
	\centering
    \includegraphics[width=0.5\textwidth]{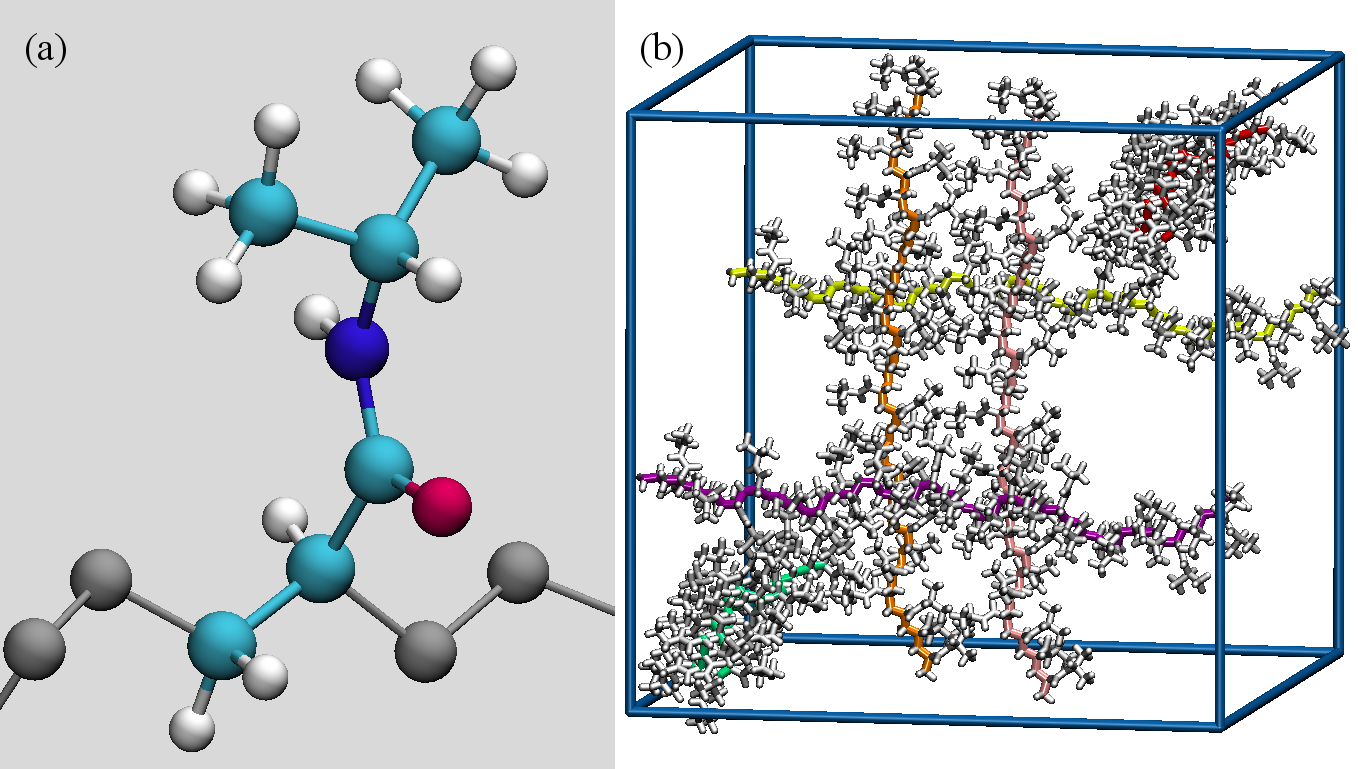}
    \caption{(a) Chemical structure of a PNIPAM repeating unit. Carbon, oxygen, nitrogen, and hydrogen atoms are shown in light blue, red, blue, and white, respectively. Backbone carbon atoms of the polymer chain are also displayed in gray. (b) Schematic representation of the atomistic model of a suspension of PNIPAM chains in water prior to density equilibration. Backbone carbon atoms of each polymer chain are represented with a different colour, while hydrogen atoms and heavy atoms in the side chain groups are displayed in gray. Water molecules are omitted for clarity.}
    \label{fig1}
\end{figure}

Poly(N-isopropylacrylamide) was purchased from Polymer Source Inc. and used without further purification. The polymer has a molecular weight $M_w=189600$ and a polydispersity index $PDI=2.88$, thus each polymer chain is composed by $\simeq$1669 repeating units. Figure~\ref{fig1}(a) reports an illustration of the chemical structure of the polymer repeating unit.

A suspension with a polymer concentration of 10~wt\% was prepared through 3 cycles of lyophilization and dispersion in D$_2$O. High concentration samples were then obtained by filling the sample holders for neutron experiments with the dispersion at 10~wt\% and then allowing the exceeding D$_2$O to evaporate at room temperature under vacuum. The concentration was regularly monitored by accurate weighing of the samples. Once reached the desired concentration of 50~wt\%, 60~wt\% and 95~wt\%, the holders were sealed and samples were left to homogenize for not less than four days. The sample with a polymer concentration of 95~wt\% is referred to as dry sample because it corresponds to the mass fraction residue determined by thermogravimetric analysis at 430~K.

\subsection{Elastic incoherent neutron scattering}
EINS experiments were carried out at the neutron spectrometers IN16B, IN13, and IN5~\cite{illwebinstr} of the Institut Laue-Langevin (ILL, Grenoble, France)~\cite{DataDOI1,DataDOI2}. The main characteristics of the instruments are summarized in Table~\ref{tab1}, whereas the details of experiments and data reduction are thoroughly described in Appendix~\ref{app:appEINS}.

The observable quantity in an EINS experiment is the intensity $I(Q,|E|\lesssim\Delta E) \doteq I(Q,0)$ scattered in a narrow energy window $\Delta E$ centered about the elastic peak $E=0$, as a function of  exchanged momentum $Q$. This observable was monitored as a function of $T$ with different energy windows $\Delta E$, as per Table~\ref{tab1}, allowing us to resolve atomic motions in a wide range of timescales $\tau$ from ps to ns. As the neutron incoherent cross-section of hydrogen exceeds by more than one order of magnitude the total (coherent plus incoherent) cross-section of deuterium and of other atomic species in the samples, the measured signal provides information on the dynamics of the polymer chains, while the water contribution is negligible, as reported in Tab.~\ref{tbl:sample}.

\begin{table}[htbp]
    \begin{ruledtabular}
        \begin{tabular}{l|cccc}
																		&IN16B		&IN13    	&IN5 (1$\Delta E$)		&IN5 (2$\Delta E$)	\\
            \hline
            $Q$-range (\AA$^{-1}$)  &0.2--1.9	&0.2--4.5	&0.55--2.0         		&0.55--2.0         	\\						
            $\Delta E$ ($\mu$eV)    &0.75  		&8    		&97     							&191          			\\
            $\tau$ (ps)   					&1800   	&150   		&15        						&7                	\\
        \end{tabular}
			\caption{$Q$-range, energy window $\Delta E$ and probed timescale $\tau$ of the different EINS measurements. For IN16B and IN13, $\Delta E$ corresponds to the instrumental energy resolution, while for IN5 it is the $E$-window used to calculate $I(Q,0)$. In the employed experimental configuration, the energy resolution of IN5 was $\sim87~\mu$eV, see Appendix~\ref{app:appEINS}.}
			\label{tab1}
    \end{ruledtabular}
\end{table}

\subsection{All-atom molecular dynamics simulations}
The \textit{in silico} model of a suspension of PNIPAM linear chains in water was designed by including in a cubic box 6 atactic polymer segments made of 30 repeating units~\cite{Tavagnacco2018}, with an extra-boundary connectivity between adjacent periodic images to mimic the effect of the higher degree of polymerization in the experimental samples, and by adding explicit water. A schematic representation of the model is shown in Fig.~\ref{fig1}(b). All-atom MD simulations were performed on linear polymer chains suspensions with PNIPAM mass fraction of 60\% using the GROMACS 5.1.4 software~\cite{Pall2015,Abraham201519}. PNIPAM and water were described using the OPLS-AA force field~\cite{Jorgensen1996} with the implementation by Siu et al.~\cite{Siu2012} and the Tip4p/ICE model~\cite{tip4pICE}, respectively. Numerical simulations were carried out in a wide range of temperatures, by cooling the system from 293 to 193~K. This allows a direct comparison with experiments. At each temperature the system was first equilibrated in a pressure bath at 1 bar, maintained by the Parrinello-Rahman barostat with a time constant of 2~ps, up to a constant density value, i.e. drift less than $2 \times 10^{-3}$~g~cm$^{-3}$ over 20~ns. Simulation data were then collected for 330~ns in the NVT ensemble, with a sampling of 0.2~frame/ps.

The leapfrog integration algorithm was employed with a time step of 2~fs. The length of bonds involving hydrogen atoms was kept fixed with the LINCS algorithm. Cubic periodic boundary conditions and minimum image convention were applied. The temperature was controlled with the velocity rescaling thermostat~\cite{Bussi2007} with a time constant of 0.1~ps. Electrostatic interactions were treated with the smooth particle-mesh Ewald method with a cutoff of non-bonded interactions of 1~nm. A total trajectory interval of about 0.5~$\mu$s was calculated for each temperature. The last 300~ns of trajectory were considered for analysis. The software VMD~\cite{VMD} was employed for graphical visualization.

\section{Results}
\label{sec:results}
\begin{figure}[htbp]
  \centering
  \includegraphics[width=0.5\textwidth]{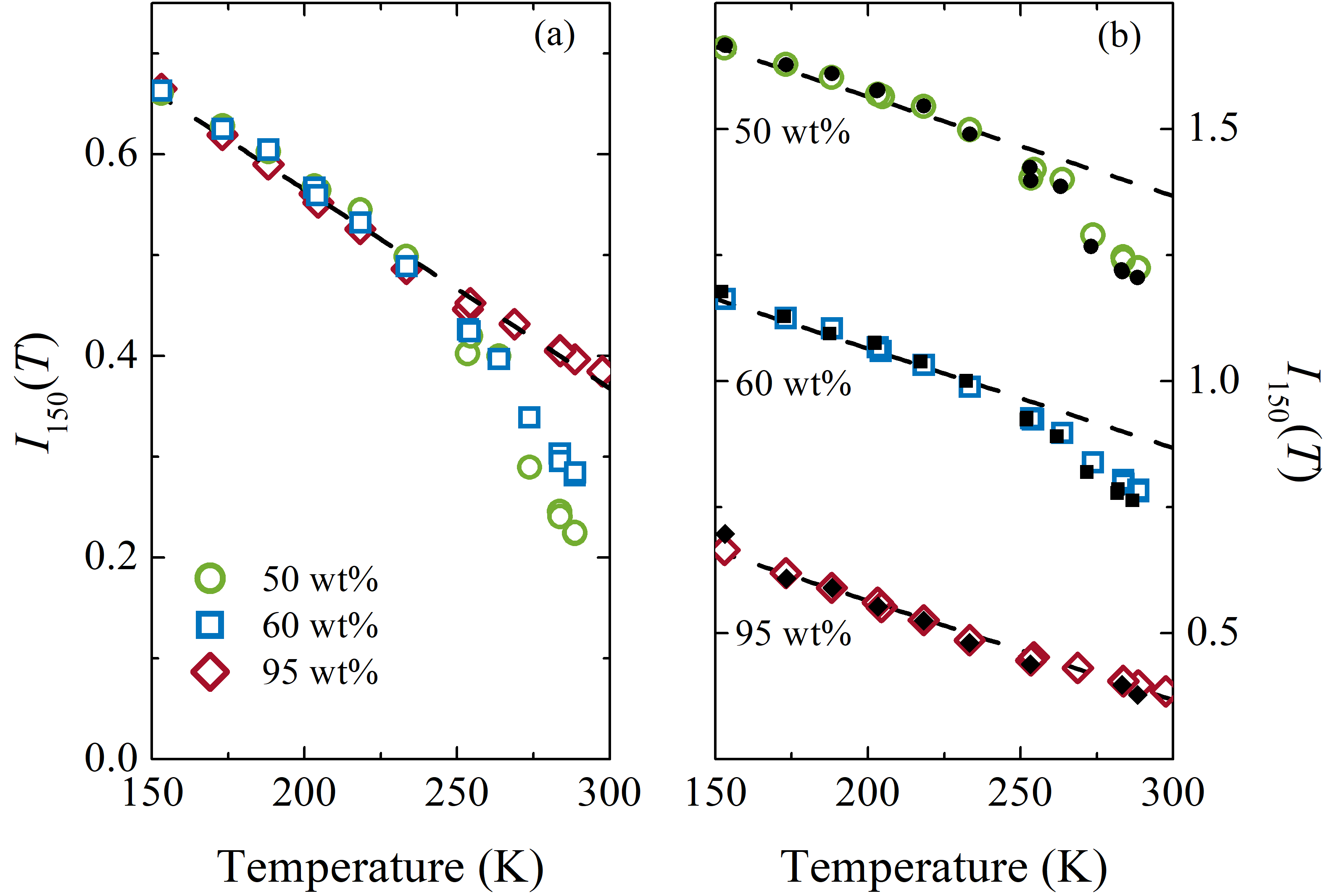}
  \caption{IN13 integrated elastic intensity $I_{150}(T)$ as a function of temperature for (a) chains as a function of PNIPAM  concentration and (b) PNIPAM chains (open symbols) compared to PNIPAM microgels (filled symbols). The dashed black lines are linear fits to the data of the dry polymer chain. Data are normalized to 1 for $T\rightarrow0$.  In (b), data at different concentrations are shifted by 0.5 for clarity. Error bars are within symbol size.}
  \label{fig2}
\end{figure}

We start by reporting the integrated elastic intensity $I_{\tau}(T)$, i.e. the integral over the whole measured $Q$-range of $I(Q,0)$ at a given $\tau$, as a function of temperature. Figure~\ref{fig2}(a) shows $I_{150}(T)$ measured at IN13 for different polymer weight fractions. While the dry system displays a linear decrease of $I_{150}(T)$ with increasing $T$, the intensity of the two hydrated samples undergoes a clear drop at a temperature that can be located roughly within 225~K and 250~K. The departure from the linear trend marks the onset of anharmonic motions ascribed to the occurrence of the dynamical transition~\cite{doster1989dynamical}. Within the available $T$-resolution, $T_d$ does not show a dependence on sample concentration, while the extent of the drop in $I_{150}(T)$ does. Hence, for the dry sample the transition disappears, in agreement with studies suggesting a driving role of water in the transition~\cite{Tarek2002,schiro2015translational,camisasca2016two}. These findings are in quantitative agreement with IN13 measurements on PNIPAM microgel networks~\cite{zanatta2018evidence}, reported for comparison in Fig.~\ref{fig2}(b). The nearly identical $T$-behavior for $I_{150}(T)$ observed for polymer chains and microgels at corresponding concentrations suggests that the mesoscopic topological features of the macromolecule, such as molecular weight and polydispersity, have negligible effects on the phenomenon. Therefore, we conclude that there is no difference between cross-linked and linear chains, highlighting the fact that the dynamical transition does not depend on the details of the internal macromolecular architecture, similarly to what observed for biological systems.

\begin{figure}[htbp]
\centering
  \includegraphics[width=0.45\textwidth]{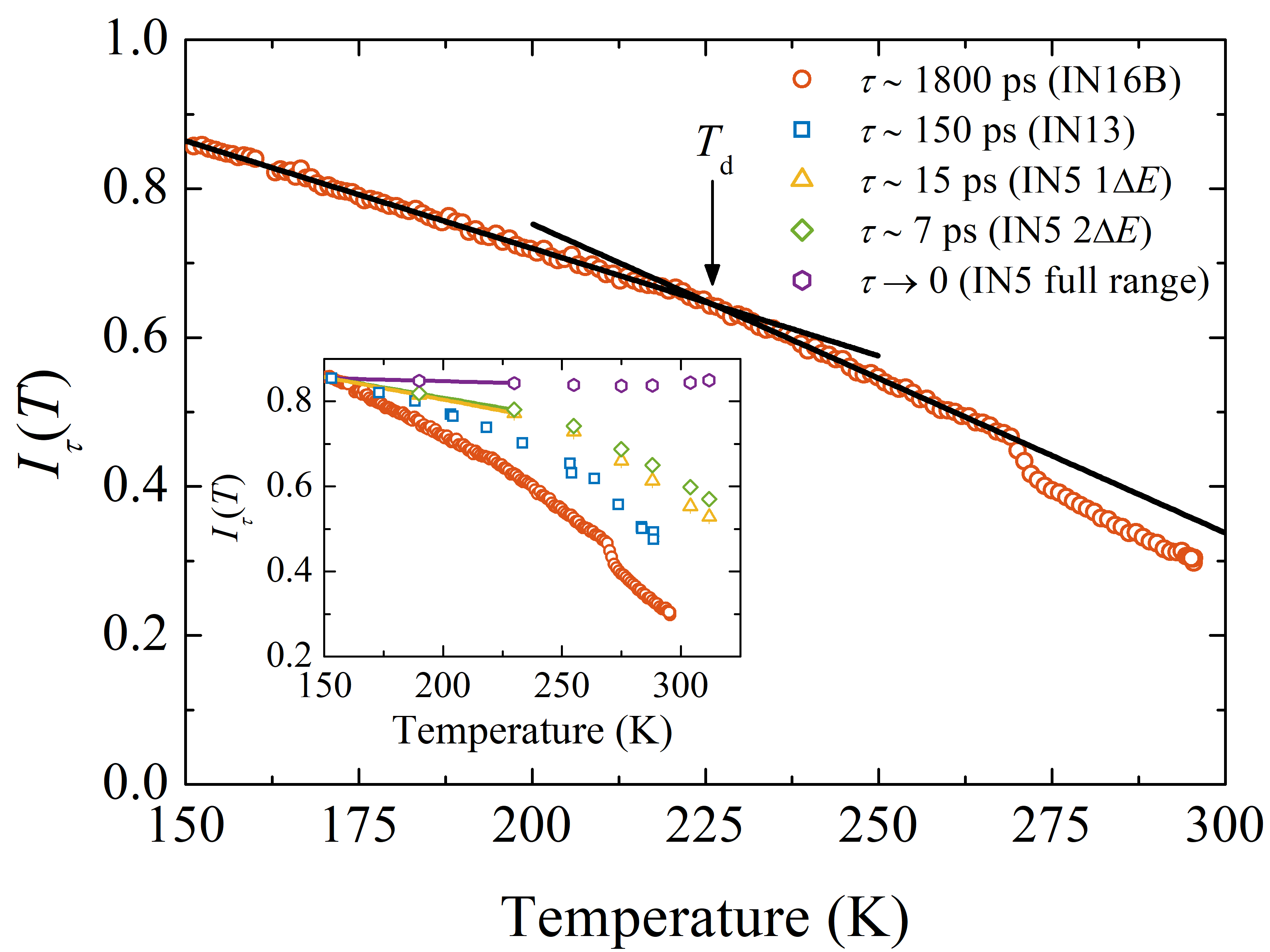}
    \caption{IN16B integrated elastic intensity of 60 wt\% linear chains. The dynamical transition temperature $T_d=226\pm1$~K is determined as the crossing point of two linear behaviours (black lines). Inset: $I_{\tau}(T)$ obtained by integrating $I(Q,0)$ measured at different $\tau$ in the common $Q$-range from 0.6 to 1.8~\AA$^{-1}$. IN13 data are normalized to IN16B ones at 150~K; for IN5 data, normalization is obtained through a linear extrapolation of the lowest-$T$ measurements (solid lines).}
  \label{fig3}
\end{figure}

Since there is no strong dependence of $T_d$ on wt\%, from now on we focus on the 60~wt\% sample only. Figure~\ref{fig3} shows the integrated elastic intensity $I_{1800}(T)$ measured at IN16B. Its higher flux allows us to obtain a finer temperature sampling than IN13, thus revealing three different regimes: (i) a linear decrease above 150~K, that in proteins is known to account for both harmonic movements and methyl groups rotation~\cite{Roh2005,schiro2010direct}; (ii) a second, steeper linear decrease above $T_d \sim 225$~K, that marks the dynamical transition; (iii) a third, sudden drop around 270~K. The latter, that is not clearly separated from (ii) on the less resolved $T$-scan performed on IN13, might be related to ice melting. To shed light on this issue, we performed complementary Differential Scanning Calorimetry (DSC) measurements (see Appendix~\ref{app:appDSC}), that confirmed the absence of macroscopic crystallization upon cooling~\cite{Afroze2000,zanatta2018evidence} and the onset of a process of cold crystallization upon heating, involving only about 2.5\% of the water molecules. As shown in Appendix~\ref{sub:EINSthermal}, this overall behavior is consistent with both cooling and heating cycles of the IN16B measurements.

The finer temperature sampling allows us also to determine with greater accuracy that the dynamical transition in linear chains takes place at about 225~K, well below the value of 250~K that was initially suggested by IN13 for microgel systems~\cite{zanatta2018evidence,tavagnacco2019water}. Given that microgels and linear chains display an identical behaviour on IN13, we expect the true $T_d$ of microgels to be close to 225~K too.

The inset of Fig.~\ref{fig3} reports a comparison among all intensities $I_{\tau}(T)$ obtained by integrating $I(Q,0)$ measured at the various $\tau$ on a common $Q$-range. Qualitatively, the data show a less and less pronounced intensity decrease as $\tau$ gets shorter, ending up in a flattening of the curve when the static approximation $\tau\rightarrow0$ is approached. All curves at finite $\tau$ exhibit a change of slope around $\sim 225$~K. Hence, the present data do not suggest a clear dependence of $T_d$ on the experimental resolution, at variance with previous reports for hydrated protein powders~\cite{schiro2012model}. This discrepancy might be due to either a true difference between PNIPAM and proteins or possibly to the sparse $T$ sampling of our IN13 and IN5 data.

For a quantitative analysis of the $T$-evolution of the polymer dynamics, we use the widely-employed double-well model  for incoherent elastic scattering~\cite{doster1989dynamical,katava2017critical}, which assumes that the sample hydrogen atoms can jump between two distinct sites of different free energy and separated by a distance $d$. Hence, $I(Q,0)$ can be written as
\begin{equation}
I(Q,0)=I_0 e^{-bQ^2}\left[1-2p_1p_2\left(1-\frac{\sin(Qd)}{Qd}\right)\right],
\label{Eq:doublewell}
\end{equation}
where $I_0$ is an intensity prefactor, $b=\left\langle\Delta u^2\right\rangle_{vib}$ is the harmonic vibrational MSD of an atom moving within a single well, $p_1$ and $p_2$ are the probabilities of finding the atom in the first or second well, respectively. For a robust determination of the physical parameters of interest, we set up a novel global analysis procedure that simultaneously fits the data acquired at all the measured timescales, over the three different $Q$-ranges and over all temperatures (see Appendix~\ref{app:fit}). Specifically, we assume that: (i) the harmonic behavior of our samples is faster than any of the experimental timescales $\tau$ in Tab. 1, thus being equally resolved by the three employed spectrometers; (ii) upon increasing $\tau$, hydrogen atoms can explore a wider spatial region, hence $d$ increases. Consequently, we constrain $<\Delta u^2>_{vib}$ to follow a linear behavior in $T$ with $\tau$-independent parameters, whereas $I_0$ and $p_1p_2$ are free to vary with both $\tau$ and $T$. In this way, we find that $d$ is roughly constant in $T$ but grows with $\tau$ as $d(\tau)=\phi\tau^{\xi}$, in agreement with previous findings for proteins~\cite{schiro2012model}. We thus constrain $d$ to follow such a power-law behavior with $\phi$ and $\xi$ as additional fit parameters. From this model, we can finally estimate the total MSD~\cite{katava2017critical} of PNIPAM hydrogen atoms for each time resolution, as
\begin{eqnarray}
\mbox{MSD}=6\left\langle\Delta u^2\right\rangle_{vib}+2p_1p_2 d^2.
\label{Eq:MSD}
\end{eqnarray}

\begin{figure}[htpb]
\centering
  \includegraphics[width=0.45\textwidth]{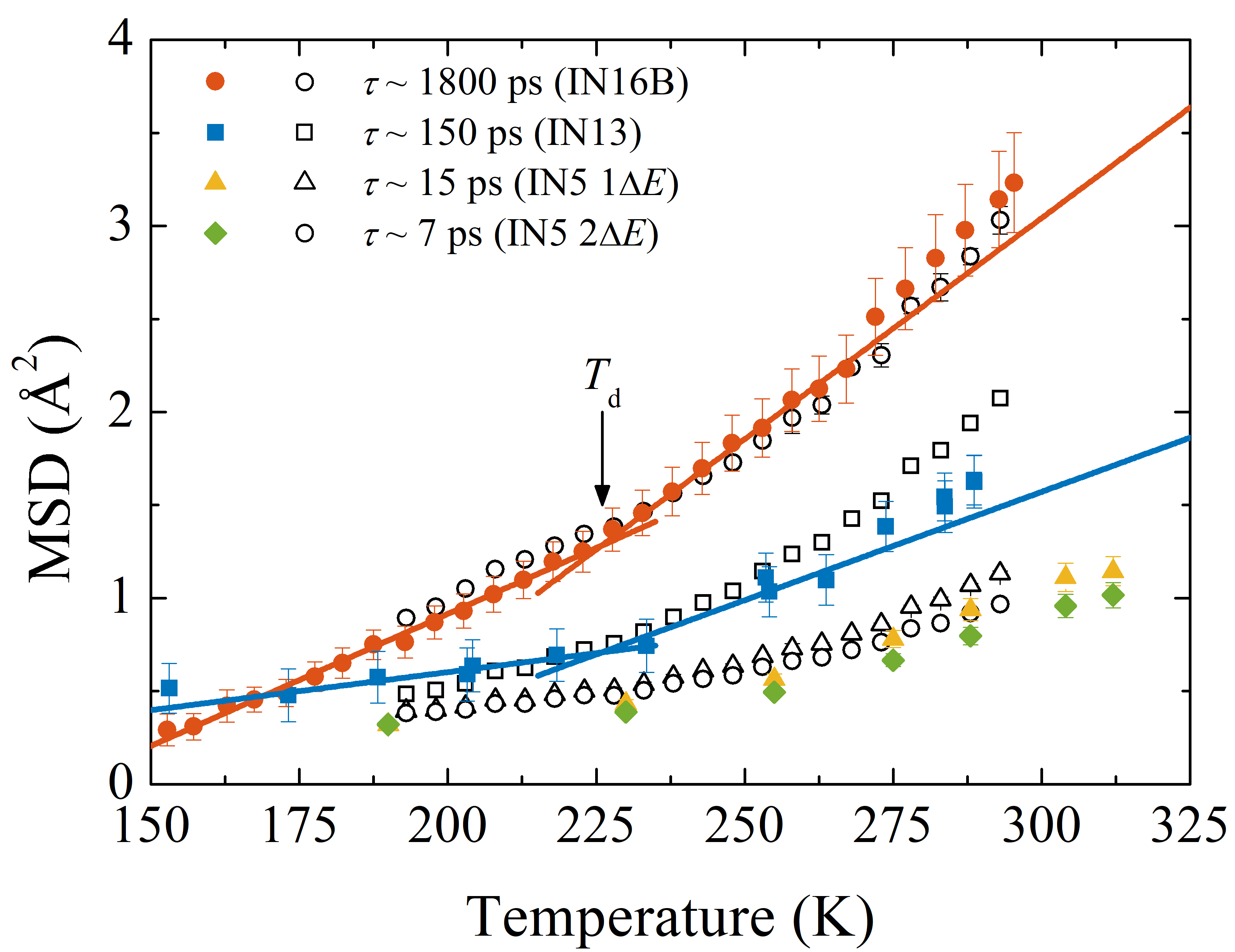}
  \caption{Comparison between experimental MSDs (filled symbols), obtained by fitting $I(Q,0)$ as described in the text, and numerical MSDs (open symbols) at the same timescales. The transition temperature obtained from the integrated elastic intensity is marked with a black arrow. Red and blue lines are guides-to-the-eye to highlight the transition in IN16B and IN13 data respectively. Error bars amount to one standard deviation and, when not visible, are within the symbol size. The high value of the first IN13 point at 150~K may be due to fluctuations in the data normalization reference, but it does not affect the change of slope of the data at higher temperatures.}
  \label{fig4}
\end{figure}

The resulting MSDs are reported in Figure~\ref{fig4}  as a function of temperature at the various experimental timescales. As expected, at a given temperature MSDs are larger for longer observation timescales $\tau$. Considering IN16B and IN13 data, we easily recognize the same three regimes also observed for the integrated elastic intensity in Fig.~\ref{fig3}. Our results do not show a clear evolution of $T_d$ with $\tau$. In proteins, the existence of this dependence is debated and results range from a strong variation of $T_d$ \cite{capaccioli2013dependence} to a substantially $\tau$-independent transition \cite{schiro2012model}. In our case, we find that $T_d\sim225$~K seems compatible with all datasets, although the sparse $T$ sampling of IN13 and IN5 does not allow to unambiguously determine a possible trend.

It is now instructive to compare the experimental results with MD simulations. The MSDs of PNIPAM hydrogen atoms directly evaluated from MD simulations at each time resolution as a function of $T$ are also reported in Fig.~\ref{fig4}. Remarkably, the numerical data quantitatively agree with the experimental estimates at all temperatures and for all measured time resolutions, without any arbitrary scaling factor. This confirms previous results obtained from IN13 measurements of microgels~\cite{zanatta2018evidence}, crucially extending their validity by more than two decades in time. In addition, the direct comparison of experiments and simulations strongly validates the global fit procedure that we have adopted to extract experimental MSDs from the measured elastic intensities. Indeed, we stress that a free fit would not be able to provide the same level of consistency both between data measured at different $E$-resolution and $Q$-range and between experimental and numerical data. Furthermore, these findings corroborate the use of the TIP4P/ICE model, which fully captures the $T$-dependence of the dynamical behavior of polymer atoms, and thus emerges as the optimal water model to simulate PNIPAM dynamics in supercooled water.

The numerical MSD data substantiate the behavior of the integrated elastic intensity and also point to the occurrence of a dynamical transition for linear PNIPAM chains around $T_d\sim$225~K. While simulations have been performed with a temperature mesh of 5~K, due to the long simulation time required for each state point, this sampling is sufficient to locate $T_d$ in good agreement with experiments. We remark that a different value of $T_d$, roughly around 250~K, was previously reported for microgels~\cite{zanatta2018evidence,tavagnacco2019water},  but  this larger value can be entirely ascribed to the broader $T$-sampling used in such earlier studies.

\begin{figure}[htbp]
\centering
  \includegraphics[width=0.5\textwidth]{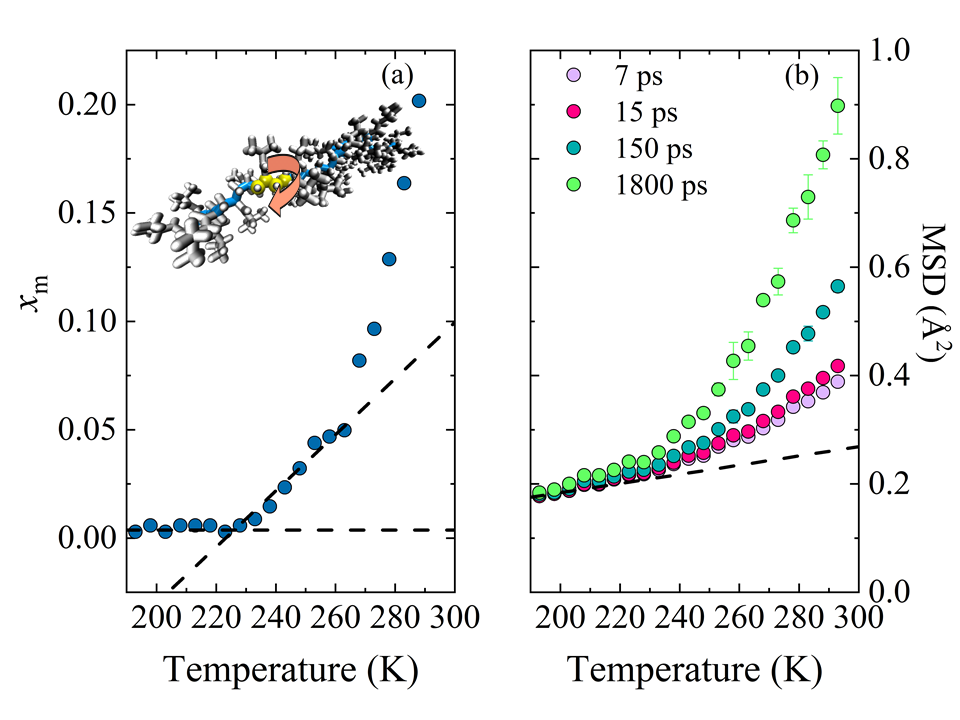}
  \caption{MD results for a) fraction of mobile backbone dihedral angles $x_m$ and b) MSD of hydrogen backbone atoms for different time resolutions, both as a function of $T$. A snapshot of a polymer chain is also shown in a), with backbone atoms in blue and a dihedral angle highlighted in yellow. In both panels, dashed lines are guides to the eye to indicate the occurrence of the dynamical transition at $T_d\sim225$~K. When not visible, error bars are within symbol size.}
  \label{fig5}
\end{figure}

The simulations further allow us to identify the specific motions underlying the onset of the transition, which should be connected, as in proteins~\cite{rasmussen1992crystalline}, to the activation of local segmental motions. This is investigated by monitoring the fraction of mobile backbone dihedral angles $x_m$ as a function of $T$, that is shown in Fig.~\ref{fig5}(a). We find that, compatibly with statistical uncertainty, $x_m$ becomes larger than zero for $T \gtrsim$ 230~K, thus confirming the molecular origin of the transition with the activation of bond flips between rotational isomeric states of the chain backbone. We stress that this analysis has been performed on the whole equilibrated run, so that its results can be considered to be independent of time resolution. Finally, it is important to note that while methyl hydrogen atoms are active at all investigated temperatures (see Appendix~\ref{app:appMD}), backbone hydrogen atoms are inactive below $T_d$, as shown by their MSDs reported as a function of $T$ in Fig.~\ref{fig5}(b) for different values of $\tau$. Clearly, deviations from a low-$T$ linear regime again occur only for temperatures higher than 225~K. These results confirm that, also in the simulation data, the value of $T_d$ is independent of the probed time window.

\section{Conclusions}
\label{sec:conclusions}
In this work we reported evidence of a low-temperature dynamical transition in linear PNIPAM polymer chains by means of EINS measurements at resolutions covering more than two decades in time. A global fit, based on the double-well model, was simultaneously applied to all measured data as a function of $T$, $Q$ and $\Delta E$, allowing us to extract the MSDs of PNIPAM hydrogen atoms. For the first time, these were found to be in quantitative agreement with those calculated in atomistic MD simulations, in the whole resolution interval and without any scaling factor. Our results clearly indicate that the dynamical transition takes place, in both linear chains and cross-linked microgels, at $T_d\sim225$~K, like in globular proteins. The value of $T_d$ seems to be independent from the experimental resolution, although further studies are needed to fully clarify this point.
We thus confirm the occurrence of a dynamical transition in hydrated powders of non-biological systems~\cite{zanatta2018evidence,iorio2019slow}, further indicating that this phenomenon is not dictated by the internal polymer architecture. It rather emerges as a general process, in those macromolecular systems that are characterized by dynamical complexity -- due to multiple stable configurations of similar energy -- and capable to efficiently confine water, couple with it by hydrogen bonding~\cite{tavagnacco2019water} and thus avoid ice crystallization. These findings further challenge the present understanding of the relationship between dynamical behavior and biological function of a biomacromolecule, and thus call for deeper investigations about the molecular mechanisms underlying the dynamical transition.

\begin{acknowledgments}
\noindent We thank Roberta Angelini and Silvia Franco for their help in performing the DSC characterization of our samples. We acknowledge ILL for beamtime and CINECA-ISCRA for computer time. LT, EC and EZ acknowledge support from European Research Council (ERC-CoG-2015, Grant No. 681597 MIMIC); LT, EB, EC and EZ from MIUR (FARE project R16XLE2X3L, SOFTART). EB, EC and EZ acknowledge support from Regione Lazio, through L.R. 13/08, Progetto Gruppo di Ricerca GELARTE n.prot.85-2017-15290.\\
LT and MZ contributed equally to this work.
\end{acknowledgments}

\appendix

\section{EINS experiments and data treatment}
\label{app:appEINS}
EINS experiments were carried out loading the samples inside flat aluminium cells ($3.0 \times 4.0$~cm) sealed with an In o-ring. The thickness of each cell was selected to achieve a transmission of about 90\% at an incoming wavelength $\lambda_i=6.271$~\AA, see Tab. \ref{tbl:sample}. The weight of each sample was checked before and after each measurement without observing any appreciable variation.

Considering the total, coherent, and incoherent scattering cross sections \cite{NeutronData}, the signal measured for each sample is largely dominated by the incoherent contribution of PNIPAM hydrogen atoms, ranging from 81 to 91\%, see Tab.\ref{tbl:sample}.

\begin{table}[htbp]
	\begin{ruledtabular}
		\begin{tabular}{l|cccccc}
			sample	& $d$		& $T$		& $\sigma_{coh}$	& $\sigma_{inc}$ 	&	$\sigma_{scatt}$ 	& $\sigma_{H}/\sigma_{scatt}$	\\
							& (mm)	&				& (b)			 			 	& (b) 		      	&	(b)				 			 	& 							      					\\
			\hline
    	50\%wt	& 0.5		& 0.89	& 163							&	826			  			& 989			 					&	0.81 \\
			60\%wt	& 0.4		& 0.89	& 128 	  				&	817			  			& 945			 					&	0.85 \\
			95\%wt	& 0.2		& 0.89	&  75							&	803			  			& 878			 					&	0.91 \\
    \end{tabular}
		\caption{Characteristics of the samples for the EINS experiment on IN16B: thickness $d$ of the cells and corresponding transmission $T$ calculated for $\lambda_i=6.271$~\AA; calculated cross sections for the coherent, incoherent and total scattering of each sample (PNIPAM and heavy water), $\sigma_{coh}$, $\sigma_{inc}$, and $\sigma_{scatt}$, respectively; ratio between the incoherent scattering cross section of PNIPAM hydrogen atoms $\sigma_{H}$ and the total scattering cross section of the sample (PNIPAM and heavy water).}		
		\label{tbl:sample}
	\end{ruledtabular}
\end{table}

\begin{figure*}[htbp]
	\centering
	\includegraphics[width=\textwidth]{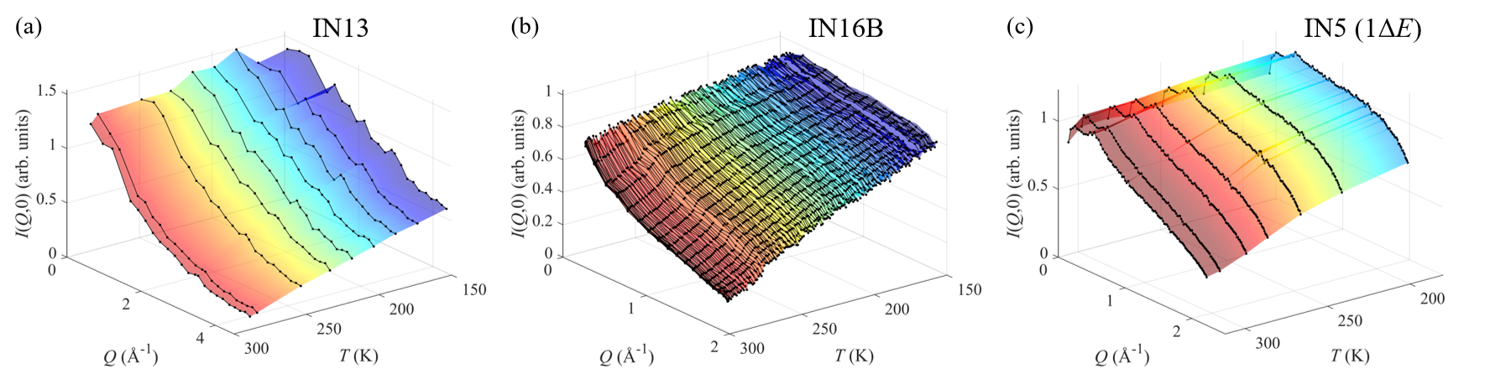}
  \caption{Temperature evolution of the elastic incoherent intensity $I(Q,0)$ for the 60 wt\% sample measured at different instruments: (a) IN16B, upon heating; (b) IN13 upon cooling; (c) IN5 upon heating and after integration over 1$\Delta E$. The color map from blue to red indicates the increasing temperature.}
  \label{figEINSdata}
\end{figure*}

\subsection{IN13 data}
\label{sub:IN13}
IN13 is a high-resolution backscattering spectrometer using thermal neutrons. In the elastic configuration, IN13 operates with an incident wavelength $\lambda_i=2.23$~\AA~and covers an interval of exchanged momentum $Q$ from about 0.2 to 4.5~\AA$^{-1}$, with an energy resolution of $8~\mu$eV, obtained as the full width at half maximum (FWHM) of a Gaussian fit to the elastic peak of a standard Vanadium sample. Measurements were carried out on the 50, 60, and 95~wt\% PNIPAM linear chain samples. The $I(Q,0)$ was acquired at selected fixed temperatures, cooling the sample from about 290~K down to 150~K and then heating back to room temperature. The acquisition time for each temperature ranged from 30 minutes to 2 hours.

Data were corrected for incident flux, cell scattering and self-shielding. The $I(Q,0)$ of each sample was normalized to a vanadium standard to account for detector efficiency fluctuations. Given the atomic composition, density and geometry of our samples, the ratio of multiple-to-single total scattering (i.e. integrated over the whole $(Q,E)$-space) can be evaluated to amount to about 16\%. In the restricted elastic window and reduced $Q$-range of our interest, multiple scattering reduces to a smaller fraction and, in mainly incoherent samples like ours, it is rather structureless as a function of $Q$. Therefore it cannot affect results such as the temperature behaviour of the data and the value of the dynamical transition temperature $T_d$. Consequently, multiple scattering corrections were neglected. Figure \ref{figEINSdata}(b) shows the so-obtained $I(Q,0)$ for the 60~wt\% PNIPAM linear chains sample.

\subsection{IN16B data}
\label{sub:IN16B}
Measurements at the high-flux backscattering spectrometer IN16B were performed in the Si(111) configuration, which produces an elastic $E$-resolution of $0.75~\mu$eV (FWHM) using neutrons with an incident wavelength $\lambda_i=6.271$~\AA. Data were acquired over a $Q$-range from about 0.2 to 1.9~\AA$^{-1}$. The $I(Q,0)$ was measured on the 60~wt\% sample during a heating ramp with a controlled heating rate of 0.3~K/min. The acquisition time was 30~s per temperature.

Data were corrected for incident flux, cell scattering and self-shielding. Each $I(Q,0)$ was normalized to a low-temperature measurement of the sample. As explained above, multiple scattering corrections were neglected. Figure \ref{figEINSdata}(a) shows the so obtained $T$-evolution of the $I(Q,0)$ for the 60~wt\% PNIPAM linear chains sample.

\subsection{IN5 data}
\label{sub:IN5}
The time-of-flight (ToF) spectrometer IN5 allows measurements of the full dynamic structure factor $S(Q,E)$ over a broad (Q,E)-range. The instrument was configured to select neutrons of incident wavelength $\lambda_i=5$~\AA, with a chopper speed of 12000~rpm. This provides an energy resolution of $87~\mu$eV (FWHM). Measurements were carried out on the 60~wt\% sample. Data were acquired at selected fixed temperatures, heating the sample from about 190~K up to 312~K. The acquisition time for each temperature was 1 hour.

The two-dimensional detector of IN5 collects the scattered neutron intensity as a function of detector pixel position (x,y), that defines the scattering angle $2\theta$, and neutron ToF. Given the isotropic nature of the sample, different (x,y) pixels corresponding to the same $2\theta$ were rebinned together into suitably spaced Debye-Scherrer cones. Data were corrected for incident flux, cell scattering and self-shielding, then normalized to a vanadium standard, and finally converted from ToF to exchanged energy, thus obtaining $I(2\theta,E)$ spectra. Once again, multiple scattering processes were neglected.

To extract the elastic intensity of interest, the $I(2\theta,E)$ spectra were integrated over two different symmetrical regions around the elastic peak, respectively with extension $1\Delta E =97~\mu$eV and $2\Delta E = 191~\mu$eV. Finally the scattering angle $2\theta$ was converted into $Q$, leading to the elastic intensities $I(Q,0)$ shown in Figure \ref{figEINSdata}(c). Due to the intersection geometry of the Debye-Scherrer cones with the detector surface, non-physical intensity drops appear at the lowest and highest $Q$-values, namely for $Q$ smaller than 0.55~\AA$^{-1}$ and larger than 2.0~\AA$^{-1}$ (see Fig. \ref{figIN5fit}). The corresponding data points were discarded during the data fitting procedure.

\section{Cold crystallization and melting}
\label{app:appDSC}

\subsection{DSC analysis}
\label{sub:appDSC}
\begin{figure}[htbp]
    \centering
  \includegraphics[width=0.4\textwidth]{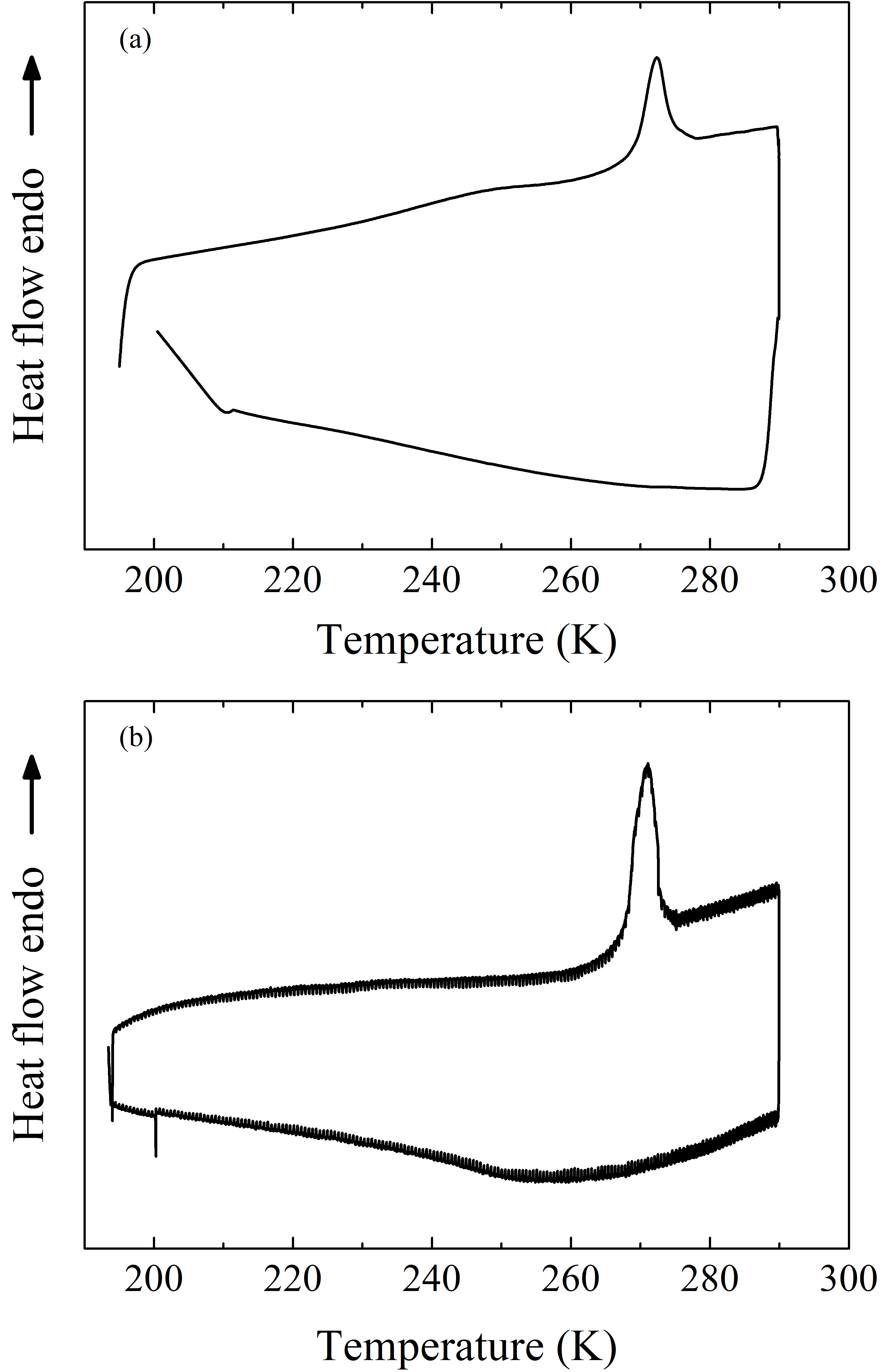}
  \caption{DSC thermograms of PNIPAM dispersions 60~wt\%, measured at (a) 10~K/min and (b) 0.3~K/min.}
  \label{FigDSC}
\end{figure}

Thermal analyses (Fig.~\ref{FigDSC}) on PNIPAM chains at 60~wt\% were recorded with a differential scanning calorimeter DSC 8000 Perkin Elmer, equipped with Intracooler II as cooling system.

DSC analyses were done on about 10-15 mg of PNIPAM dispersion at a concentration of 60~wt\% in D$_2$O. Measurements were carried out under nitrogen atmosphere (20~mL/min) in a sealed pan of aluminium to prevent changes in concentration during the heating/cooling steps. Samples were prepared starting from the material used in EINS experiments. PNIPAM linear chains were dried up and then dispersed again in D$_2$O to obtain a concentration of 10~wt\%. The target concentration was reached by evaporating the exceeding D$_2$O, then pans were sealed and analysed.

The measurement shown in Figure~\ref{FigDSC}(a) was carried out by cooling the system from 298 to 193~K at 40~K/min, then heating it to 290 K and finally by cooling it again to 193 K with a scanning rate of 10~K/min. The measurement reported in Figure~\ref{FigDSC}(b) was carried out by cooling the system from 298 to 193~K with a scanning rate of 40~K/min, then heating back to 290~K and finally cooling again to 193~K with a scanning rate of 0.3~K/min. The first cooling steps at 40~K/min are not reported here. The second protocol was followed to simulate the thermal history of the samples during the EINS experiments.

Overall the two thermograms obtained at different scanning rate are very similar, with the same $T_{peak}$ at 271~K. In both cases, no crystallization peaks were detected under cooling, even at the lower scanning rate. However, during the heating step, a process of cold-crystallization (exothermic peak) followed by fusion (endothermic peak) was observed, thus indicating the presence of a small amount of crystallized D$_2$O,  following glass melting upon re-heating. The values of $T_{onset}$, $T_{peak}$ and the melting enthalpy ($\Delta H_m$), extracted from the onset, the maximum of the peak and the area of the peak, respectively, are reported in Table~\ref{tbl:tableDSC}.

\begin{table}[htbp]
	\begin{ruledtabular}
    \begin{tabular}{ccccc}
			scan rate & $T_{onset}$ & $T_{peak}$ 	& $\Delta H_m$ 	& $X_c$ \\
			(K/min)	  & (K)					& (K) 				& (J/g) 				& (\%) 	\\
			\hline
			10				& 268.7				& 272.3				& 7.54					& 0.9		\\
			0.3				& 270.7				& 271.1				& 20.89					& 2.5		\\
		\end{tabular}
		\caption{Onset ($T_{onset}$) and melting temperature ($T_{peak}$), melting enthalpy ($\Delta H_m$) of D$_2$O and degree of crystallinity ($X_c$) in PNIPAM 60~wt\% for the two different scanning rate.}
		\label{tbl:tableDSC}
	\end{ruledtabular}
\end{table}

The peak at $\sim$270~K confirms the hypothesis that the sudden drop observed in the integrated elastic intensity $I_{1800}(T)$ measured at IN16B is related to ice melting. The degree of crystallinity $X_c$ in the dispersions can be calculated, using the enthalpy obtained from the peak area, with the following equation:
\begin{equation}
X_c~(\%)=\frac{\Delta H_m}{\Delta H^{0}_m}\cdot \mbox{D}_2\mbox{O~(wt\%)},
\label{Eq:cristallinity}
\end{equation}
\noindent where $\Delta H^{0}_m$ is the standard fusion enthalpy of deuterium oxide at 276.7~K, that is 340.7~J/g.
The degree of crystallinity is reported in Tab. \ref{tbl:tableDSC} and shows a crystallization of only a very small fraction of heavy water. It is worth noting that $X_c$ measured in (b) is slightly higher, probably due to the lower scanning rate.

\subsection{Comparison between cooling and heating scans in EINS experiments}
\label{sub:EINSthermal}
\begin{figure}[htbp]
	\centering
  \includegraphics[width=0.45\textwidth]{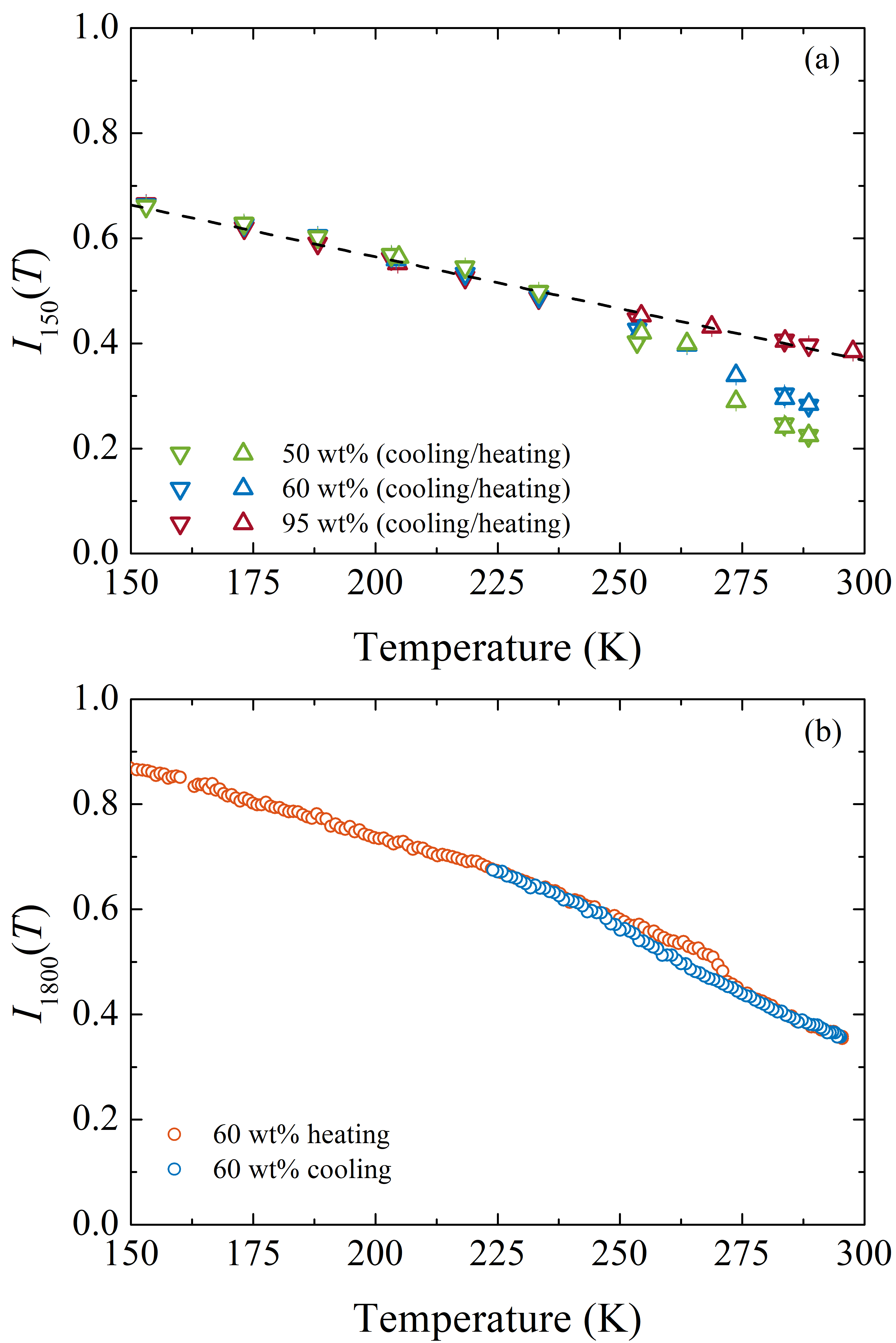}
  \caption{Comparison between the $Q$-integrated intensities acquired during cooling and heating cycles, on IN13 (a) and IN16B (b).}
  \label{FigEINStherm}
\end{figure}

The IN13 and IN16B data were measured  both upon cooling and upon heating. As described in \ref{sub:IN16B} and \ref{sub:IN13}, on IN13 the $I(Q,0)$ was acquired in static mode at fixed temperature steps for 1 hour each, whereas on IN16B the $I(Q,0)$ was saved every 30 seconds along a dynamic temperature scan with constant heating or cooling rate. The resulting integrated intensities are compared in Fig.~\ref{FigEINStherm}. While on IN13, Fig.~\ref{FigEINStherm}(a), there is no difference between cooling and heating cycles, a small but clear hysteresis appears in the IN16B data, Fig.~\ref{FigEINStherm}(b).

The absence of hysteresis on IN13 is due to both the much longer acquisition time and the static temperature mode adopted for the measurements. Both experimental choices are imposed by the lower neutron flux of IN13. As a consequence, the sample is allowed for long equilibration times at each measured temperature and is always in thermodynamic equilibrium, which results in perfectly overlapping heating and cooling data points. This is not the case for the IN16B data, where the temperature ramp mode with very fast acquisition times reveals the effects of possible temperature gradients in the sample and the details of its thermodynamic evolution. Indeed, a sudden intensity drop is observed in the heating cycle around 271~K, which is not present along the cooling cycle. The heating and cooling curves merge again around 240~K.

The hysteresis observed for IN16B data is actually in very good agreement with the calorimetric data reported in Fig.~\ref{FigDSC}. Indeed, it is clear that the intensity drop at 271~K, witnessing a sudden increase of atomic mobility, is consistent with the sharp peak in the heating DSC ramp due to the fusion of the small amount of cold-crystallized water reported in Tab.~\ref{tbl:tableDSC}. Instead, the smooth behaviour of EINS data upon cooling is reflected by the cooling DSC ramp where no crystallization peaks are present.

\section{EINS data analysis and fit procedure}
\label{app:fit}
As thoroughly described in the Sec.~\ref{sec:results}, EINS data were fitted using the double-well model \cite{doster1989dynamical,katava2017critical}. Here we simply recall the analytical expression foreseen by the model for $I(Q,0)$:
\begin{equation}
I(Q,0)=I_0 e^{-bQ^2}\left[1-2p_1p_2\left(1-\frac{\sin(Qd)}{Qd}\right)\right],
\label{Eq:doublewellA}
\end{equation}
where $I_0$ is an intensity prefactor, $b=\left\langle\Delta u^2\right\rangle_{vib}$ is the harmonic vibrational MSD of an atom moving within a single well, $p_1$ and $p_2$ are the probabilities of finding the atom in the first or second well, respectively.

\begin{figure}[htbp]
	\centering	
	\includegraphics[width=0.45\textwidth]{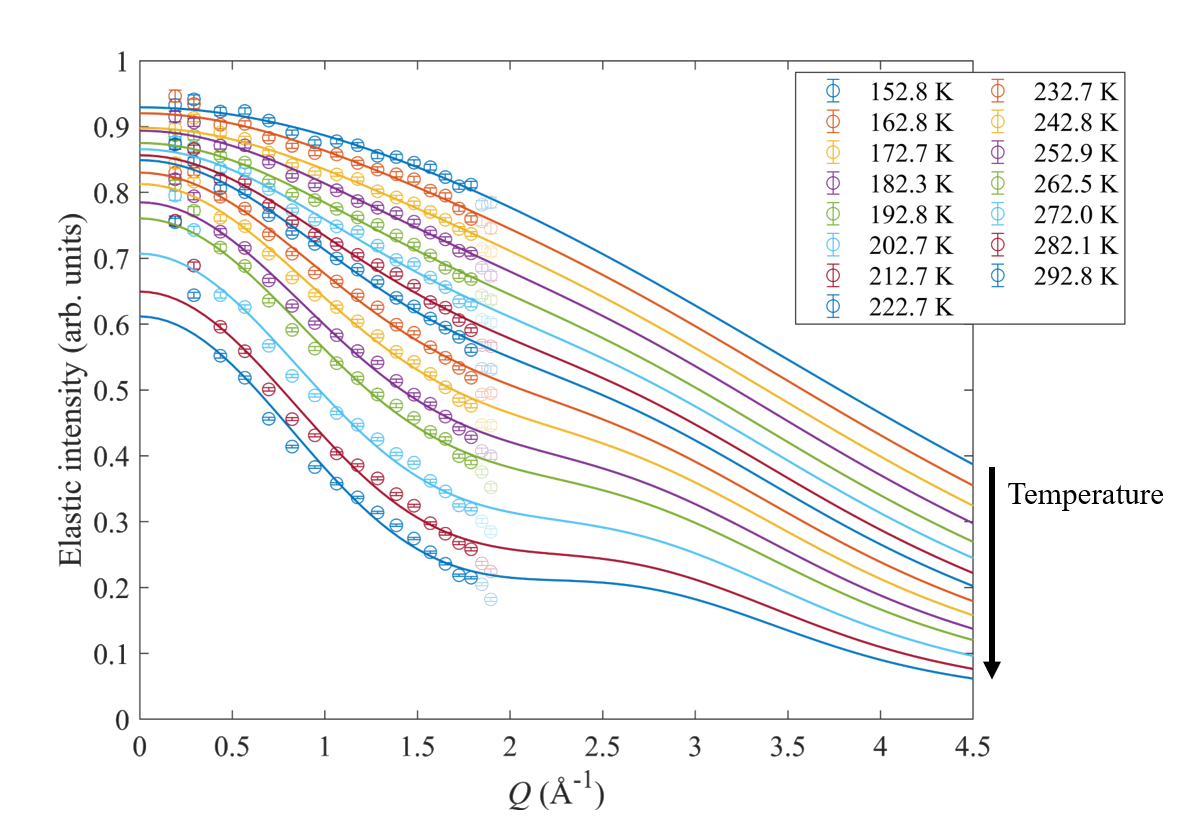}
  \caption{Typical fits of the $I(Q,0)$ measured on IN16B in PNIPAM linear chains at 60~wt\%. For sake of clarity, only selected temperature are reported. Shaded data were not used in the fit procedure. Colors are as in the legend and the black arrow indicates the direction of the increasing temperatures.}
  \label{figIN16Bfit}
\end{figure}

\begin{figure}[htbp]
	\centering
  \includegraphics[width=0.45\textwidth]{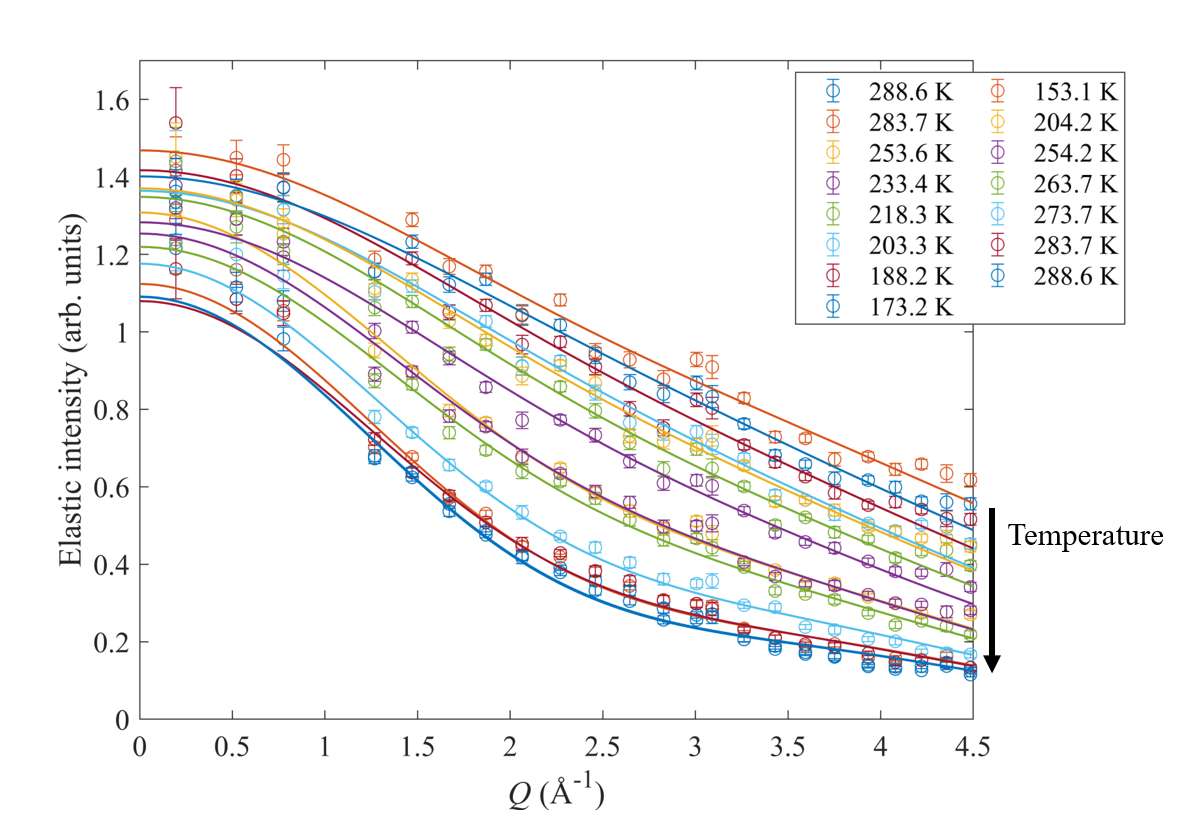}
  \caption{Typical fits of the $I(Q,0)$ measured on IN13 in PNIPAM linear chains at 60~wt\%. Colors are as in the legend and the black arrow indicates the direction of the increasing temperatures.}
  \label{figIN13fit}
\end{figure}

\begin{figure}[htbp]
	\centering
  \includegraphics[width=0.45\textwidth]{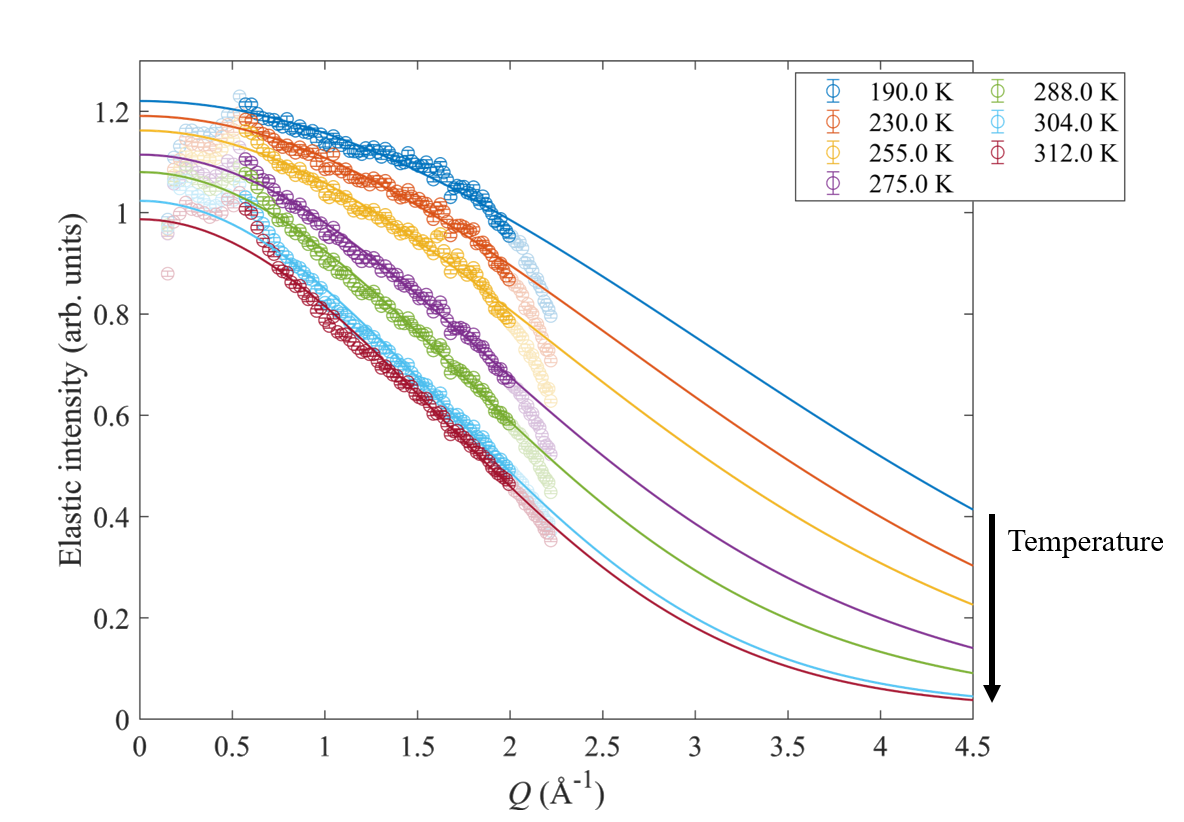}
  \caption{Typical fits of the $I(Q,0)$ measured on IN5 in PNIPAM linear chains at 60~wt\%. Shaded data were not used in the fit procedure. Colors are as in the legend and the black arrow indicates the direction of the increasing temperatures.}
  \label{figIN5fit}
\end{figure}

The data acquired at the four different energy resolutions, over the three different $Q$-ranges and over all temperatures were fitted simultaneously within a global fit procedure. To do so, we assumed that:
\begin{itemize}
	\item $I_0$ depends both on $\tau$ and $T$, so it is a \textit{local} parameter for each $I(Q,0)$;
	\item $b$ does not depend on $\tau$, while following a linear dependence on $T$, then $b(T)=q+mT$ with $q$ and $m$ \textit{global} parameters common to all the datasets;
      \item $p_1$ and $p_2$ depend both on $\tau$ and $T$ so they are \textit{local} parameters for each $I(Q,0)$;
	\item $d$ does not depend on $T$;
	\item the dependence of $d$ on $\tau$ can be written as a power law of the form $d=\phi\tau^{\xi}$ \cite{schiro2012model}, where $\phi$ is related to the diffusion coefficient, while the exponent $\xi<0.5$ takes into account a possible subdiffusive behaviour often observed in polymeric systems \cite{weber1993}.
\end{itemize}
Results for the 60~wt\% sample are shown in Figs. \ref{figIN16Bfit}, \ref{figIN13fit} and \ref{figIN5fit}. Within this approach, the model describes very well our data, providing also a good agreement with the simulations, see Fig.~\ref{fig4}.

To improve the signal to noise ratio, the $I(Q,0)$ acquired on IN16B were binned over $T$-channels of 5~K. The temperature of the final data is the average temperature of the $I(Q,0)$ inside each $T$-channel.

\section{PNIPAM internal dynamics from Molecular Dynamics simulations}
\label{app:appMD}
We investigate the onset of anharmonic motions in the polymer chains by monitoring the conformation and torsional dynamics of dihedral angles. In the analysis of the torsional dynamics of methyl groups in PNIPAM side chains, we defined the dihedral of the methyl group as the angle formed by the atoms $\mbox{N}-\mbox{C}_{isopropyl}-\mbox{C}_{methyl}-\mbox{H}$, whereas for backbone dihedral angles we considered four consecutive backbone carbon atoms. As shown in Table~\ref{tbl:dihedral}, transitions between conformational states of methyl groups are observed at each temperature and all the dihedrals angles are active in the whole temperature range. On the contrary, in the case of backbone dihedral angles, a clear increase of the number of mobile dihedral angles occurs at $T_d$.

\begin{table}[htbp]
	\begin{ruledtabular}
			\begin{tabular}{c|cc||c|cc}
        $T$~(K)	&	\multicolumn{2}{c||}{\textbf{$x_m$}}&$T$~(K)	&	\multicolumn{2}{c}{\textbf{$x_m$}} \\
								& $Backbone$ & $Methyl$	&	& $Backbone$ & $Methyl$							\\
				\hline
				193 & 0.29 & 100 &243 & 2.3 & 100 \\
				198 & 0.58 & 100 &248 & 3.2 & 100 \\
				203 & 0.29 & 100 &253 & 4.4 & 100 \\
				208 & 0.58 & 100 &258 & 4.7 & 100 \\
				213 & 0.58 & 100 &263 & 5.0 & 100 \\
				218 & 0.58 & 100 &268 & 8.2 & 100 \\
				223 & 0.29 & 100 &273 & 9.6 & 100 \\
				228 & 0.58 & 100 &278 & 13  & 100 \\
				233 & 0.88 & 100 &283 & 16  & 100 \\
				238 & 1.5  & 100 &288 & 20  & 100 \\
			\end{tabular}
			\caption{Torsional dynamics of the polymer dihedral angles. $T$ is the temperature and $x_m$ is the percentage of mobile dihedrals of the backbone and the methyl groups. Analysis over the last 300~ns.}
			\label{tbl:dihedral}
	\end{ruledtabular}
\end{table}
We quantitatively compared the experimental and numerical results by calculating the numerical MSD of PNIPAM hydrogen atoms from the following equation:
\begin{eqnarray}
\mbox{MSD}(t)&=&\langle |r_H(t)-r_H(0)|^2 \rangle
\label{Eq:MSDsim}
\end{eqnarray}
where $r_H(t)$ and $r_H(0)$ are the position vectors of a PNIPAM hydrogen atom at time $t$ and $0$, averaged over time origins and hydrogen atoms.

\begin{figure}[htbp]
	\centering
    \includegraphics[width=0.4\textwidth]{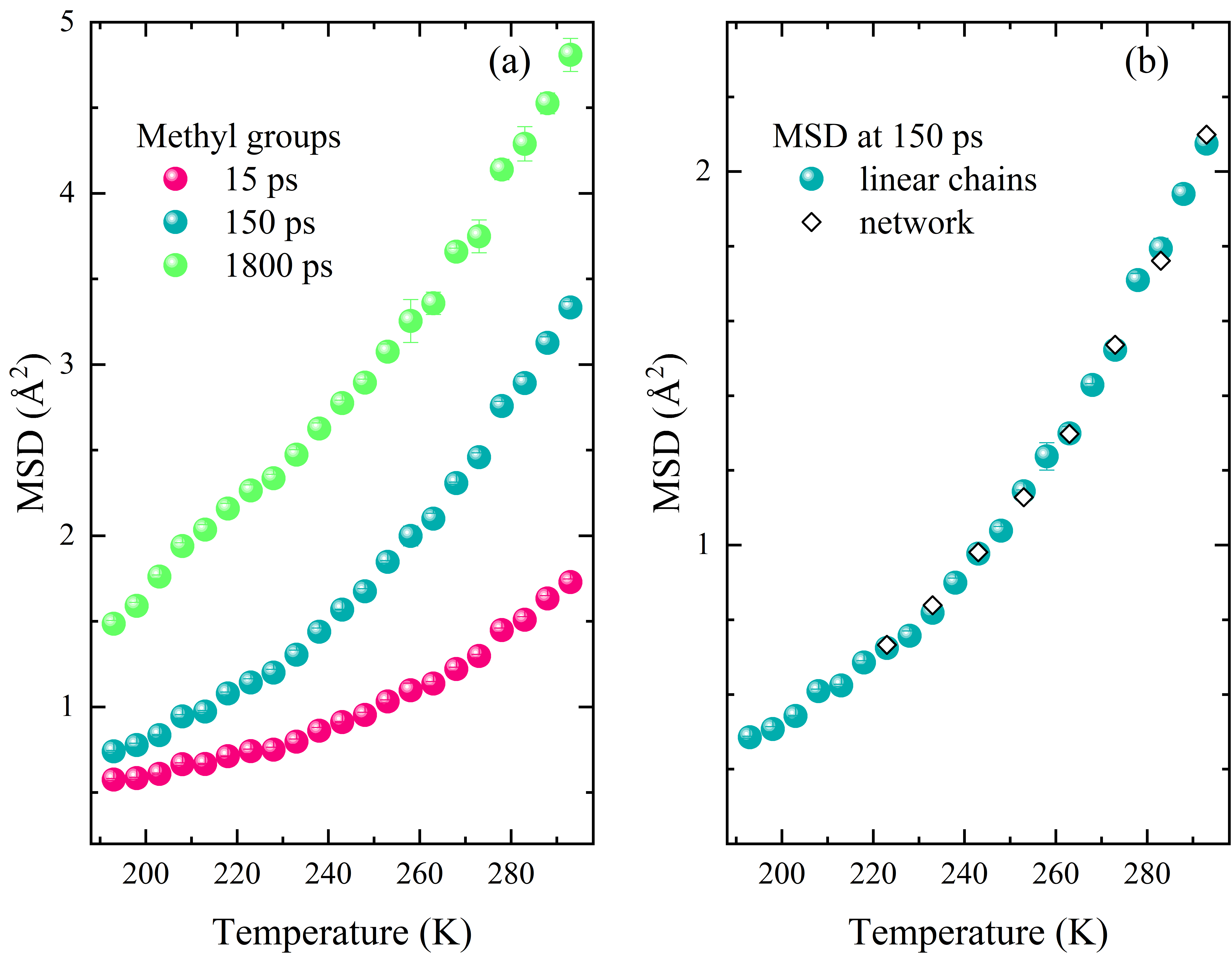}
    \caption{(a) Temperature dependence of MSDs of hydrogen atoms belonging to the methyl groups calculated at different time resolution: 15~ps (pink), 150~ps (blue) and 1800~ps (green). (b) Comparison between the temperature dependence of PNIPAM hydrogen atoms MSDs calculated from MD simulations at 150~ps at a concentration of 60~wt\% of linear polymer chains (blue circles) or polymer network (white diamonds)~\cite{tavagnacco2019water}. When not visible, error bars are within symbol size.}
    \label{FigMSDSIM}
\end{figure}

In addition to the MSDs of all PNIPAM hydrogen atoms and those in the backbone, we have also calculated the contribution of the MSDs of the hydrogen atoms in the methyl groups which also exhibits a change at $T_d \sim 225$~K, see Fig.~\ref{FigMSDSIM}(a).

Finally, to evaluate the role of the molecular architecture on the polymer local dynamics, we have reported in Fig.~\ref{FigMSDSIM}(b) a comparison between the temperature dependence of the MSDs of PNIPAM hydrogen atoms of linear chains and microgels network, whose data were taken from Ref.~\cite{tavagnacco2019water}. Fig.~\ref{FigMSDSIM}(b) shows a quantitative agreement between the MSDs calculated at 150~ps with no differences between the two polymer architectures.

%

\end{document}